\documentstyle[epsfig,subfigure]{jpsj}
\title{The Influence of Percolation in the generalized 
Chalker--Coddington Model}
\author{Marcus {\sc Metzler}}
\inst{Department of Physics, Toho University, Miyama 2-2-1, Funabashi, 
Chiba 274-8510}
\abst{
We numerically investigate the influence of classical percolation on
the quantum Hall localization-delocalization transition. This is accomplished
within the framework of the generalized Chalker--Coddington network
model which allows us to control the number of {\em classical} saddle
points by setting the width $W$ of the saddle point distribution. It
is found that increasing this width causes a new microscopic length
scale to appear which depends on $W$ and scales with the exponent
$X\approx 1.36$  which indicates a close connection to the classical
percolation length $\xi$ and its exponent
$\nu_p=4/3$. Furthermore, the influence of an increase in $W$ 
on the spectral statistics of the quasienergies of the network model
is investigated. An effect similar to the increase of the potential
correlation length in the Landau model is seen.
} 
\kword{
Chalker-Coddington model, level spacing distribution, scaling, 
quantum Hall systems, multifractality}

\twocolumn
\begin{document}
\maketitle
\section{Introduction}
The Chalker-Coddington network \cite{Chalker_Coddington}
is a model for quantum Hall systems with
long ranged disorder potentials. It represents a system of
two-dimensional (2D) electrons in a strong magnetic field and smooth
disorder potential. As a member of the quantum Hall universality
class \cite{book} it has been used to determine various critical 
quantities at the
localization-delocalization (LD) transition point between the quantized
plateaus of the Hall conductance
\cite{Chalker_Coddington,Eastmond,LWK,Klesse_Metzler1}. 

The model is based on the semi-classical time evolution picture of 2D
electrons moving along the equipotential contours of a smooth disorder
potential under the influence of a strong magnetic field. The electronic
states are defined by the amplitudes on the network links representing
the equipotential contours and the time
evolution is determined by unitary scattering matrices at the nodes
of the network corresponding to the saddle points
of the random potential  where tunneling between contours occurs.
The scattering strength is determined by the electron energy and the energy
of the saddle point. In contrast to the original network model introduced by 
Chalker and Coddington,
who explicitly excluded any percolation effects by setting all saddle point
energies to zero, the generalized version of the model allows percolation
effects by introducing an energy range $[-W,W]$ for the distribution of 
saddle point energies \cite{Eastmond,LWK,Klesse_Metzler1}. 
While this generalization of the model
does not change the critical behavior as long as the investigated systems
are large enough, it introduces a microscopic length scale $a$ that depends on 
$W/E_t$, where $E_t$ is the tunneling energy at the saddle points 
\cite{Klesse_Metzler1}. Since the motion of electrons with energy $E$ at 
saddle points with energies $u_k$ that obey $|u_k-E|\gg E_{\rm t}$
follows the classical path,
it was concluded in refs.~\citen{Klesse_Metzler1} and \citen{Polyakov} 
that this length scale must be connected to the classical percolation 
length, meaning that $a$ scales with the classical percolation exponent 
$\nu_p=4/3$, i.e. 
\begin{equation}
  \label{eq:aofW}
a\left(\frac{W}{E_t}\right)\propto \left(\frac{W}{E_t}\right)^{\nu_p}.
\end{equation}
As long as the system size $L$ is much larger than this length scale the
generalized model will show the same critical properties as the original.

In the following we will first give a short description of the generalized
Chalker--Coddington model followed by a review of the arguments given
in refs.~\citen{Klesse_Metzler1} and \citen{Polyakov} for the
influence of percolation 
effects. We will then give numerical evidence that the microscopic length 
scale indeed shows the scaling behavior we expected. After that we take a
look at the spectral properties of the network model and their dependence
on $W$. We will show that the shape of the level spacing distribution
function at criticality changes more and more to Poissonian behavior 
with increasing $W$. This is also the case for the number variance and 
consequently the spectral compressibility. In spite of these changes
the scaling exponent $\nu$ seems to be unaffected. The change of shape 
in the level spacing distribution is similar to that observed by Ono
et al. \cite{OOK} when increasing the correlation length of the
disorder potential.

\section{The Network Model}
As we have already mentioned the Chalker--Coddington model is based on the 
semi-classical picture of electrons in two dimensions moving under the 
influence of a strong perpendicular magnetic field $B$ in a long-ranged
disorder potential $V$, i.e. the correlation length $l_V$ of $V$ is large 
compared to the magnetic length $l_c=\sqrt{\hbar c/eB}$. 
Starting from this picture the following network model was developed
\cite{Chalker_Coddington,Fertig}. 
\begin{figure}
  \begin{center}
    \epsfig{figure=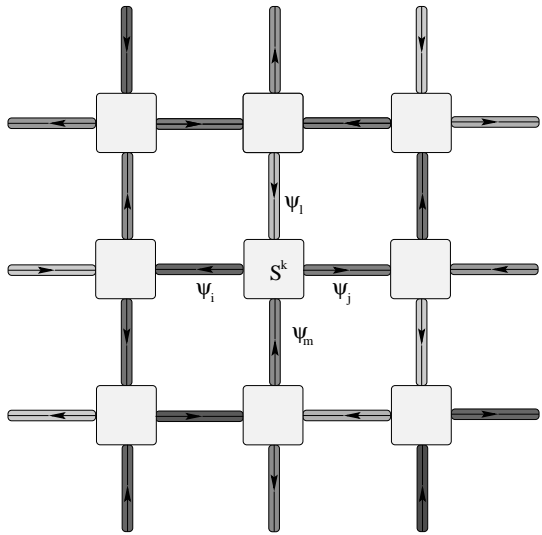,angle=270,width=0.4\textwidth}
  \end{center}
  \caption{The Chalker-Coddington network model. At each saddle point
    the matrix $S^k$ describes the scattering from incoming to
    outgoing channels. The network operator $U$ performs the same task
    for the entire network.}
    \label{fig:CC}
\end{figure}

The network consists of a 2D regular lattice (see Fig.~\ref{fig:CC}) whose links
are unidirectional channels and whose nodes are scattering centers represented
by $2\times 2$ unitary scattering matrices $S^k$, where $k$ is the node index.
The matrices $S^k$ map the amplitudes of the two incoming links onto the two
outgoing ones. The unidirectional links represent the equipotential contours 
of the potential $V$ and their random length is simulated by a random kinetic
phase given to each link. The nodes represent the saddle points of $V$ where
the electrons can tunnel between contours. The tunneling  amplitude $T$ for
each saddle point is given by $T=(1+\exp((E-u_k)/E_t))$\cite{Fertig}, 
where $u_k$ is the
energy of the saddle point with index $k$ and $E_t$ is the
tunneling energy. $E_t$ is of the form $E_t=\frac{l_c^2}{2\pi}c$, where
$c$ is the average curvature of the random potential which can be approximated
by $c\approx V_0/l_V^2$ with $V_0=\langle V^2\rangle^{1/2}$. Taking
into account that the saddle point energies are randomly distributed
in the interval
$[-W, W]$ gives us the following approximation for the tunneling 
energy\cite{Klesse_Metzler1}:
\begin{equation}
  \label{eq:tunnel}
  E_t=W\left(\frac{l_c}{l_V}\right)^2.
\end{equation}
This relation connects the quotient $E_t/W$ which is the relevant
parameter of the network model directly to the relevant
parameter $l_c/l_V$ of other smooth disorder models. It also connects
the two limiting cases of the network model ($W/E_t=0$ the original
Chalker--Coddington case and $W/E_t=\infty$ the classical network percolation
case) to the uncorrelated potential ($l_V$=0) and the classical motion 
in a magnetic field and random potential ($l_c=0$) case.

The system shows an LD phase transition when $E$ approaches the critical
energy $E_c=\langle V\rangle=0$. In this case the correlation length $\xi$ scales as
$\xi\propto |E-E_c|^\nu$, where $\nu\approx2.3$. For large enough systems 
the correlation length exponent $\nu$ does not depend on $W$\cite{LWK}.
This is also true for other critical exponents like $\alpha_0\approx2.27$ 
determined by multifractal analysis of critical 
eigenfunctions\cite{Klesse_Metzler1}.

A wave function or state of the network is defined by a 
normalized vector $\Psi=(\psi_1,\dots,\psi_n)$ whose elements $\psi_i$ are
the complex amplitudes on the $n=2L^2$ network links, where $L$ is the
system size. A wave function that at all
nodes $k$ of the network
obeys the scattering condition $\psi_{l(i,k)}=\sum_j S^k_{ij}\psi_{l(j,k)}$, 
where $l(i,k)$ maps the matrix index $i\in\{0,1,2,3\}$ of the scattering 
matrix $S^k$ at node $k$ to the respective link of the network, is stationary 
under scattering and therefore an eigenfunction of the system.

We can take all
the matrix elements of the $S^k$ and arrange them into a single operator $U$,
so that the stationarity condition becomes
\begin{equation}
  \label{eq:Uscat}
  U(E)\Psi=\Psi.
\end{equation}
The network operator $U$ also functions as a time evolution operator for the 
network states, i.e. $\Psi(t+\tau)=U\Psi(t)$, where $\tau$ is the  
characteristic scattering time\cite{Klesse_Metzler3,Klesse_PhD}.
Equation~(\ref{eq:Uscat}) states that a wave function which is an eigenfunction of 
$U(E)$ with eigenvalue 1 is an eigenfunction of the modeled system. 
Such eigenfunctions will only occur at discrete values $E_n=E$ forming the 
eigenenergy spectrum of the system. These eigenenergies are not easily 
determined, but it was found in ref.~\citen{Klesse_Metzler2} that the 
eigenvalues $\omega_\alpha$ defined by the equation
\begin{equation}
  \label{eq:Ueigen}
  U(E)\Psi_\alpha=e^{i\omega_\alpha(E)}\Psi_\alpha
\end{equation}
for a fixed value of $E$ show the same statistics as the eigenenergies $E_n$
close to $E$. These so called quasienergies $\omega_\alpha(E)$ are much easier
to determine. This and the fact that one can choose the exact point on the 
energy scale where we want to investigate the statistics make them the ideal
tool for spectral investigations\cite{Klesse_Metzler2,Metzler_Varga}. We can set $E=E_c$ and
use all the critical quasienergies $\omega_\alpha(E_c)$, i.e. the eigenphases
of $U(E_c)$ to determine spectral properties at criticality. This is an
enormous advantage compared to other methods where only a small fraction of the
spectrum is critical. We can even improve this method if we use the fact that
for every eigenphase $\omega_\alpha$ of the network operator the phase
$\omega_\alpha+\pi$  (or in the case of double periodic boundary conditions
the phases $\omega_\alpha+\pi/2$, $\omega_\alpha+\pi$ and 
$\omega_\alpha+3\pi/2$) is also an eigenphase (see Appendix~\ref{appendix}).
This leads to a reduction of the matrix size increasing the speed of the 
numerical determination and also the attainable system sizes.
Additionally, every eigenstate determined for a unitary operator $U(E_c)$ at 
the critical energy $E_c$ is critical and can be used for multifractal 
analysis\cite{Klesse_Metzler1,Klesse_Metzler3}.

\section{Percolation and Multifractality}
\label{sec:percolation}
\begin{figure}[htbp]
  \begin{center}
    \leavevmode
    \epsfig{figure=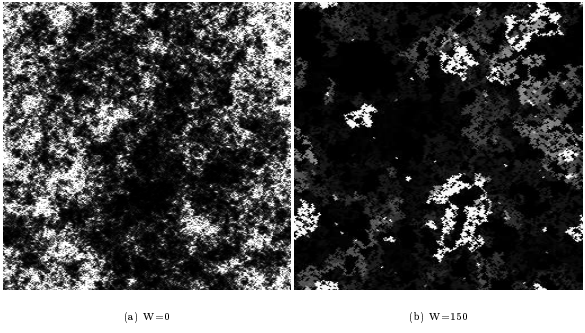,width=.45\textwidth}
  \caption{Critical wave functions 
    ($E=0$) for $W=0$ and $W=150$. The system 
    size is 256x256 saddle points. Darker areas denote lower square amplitude.}
    \label{fig:wave}
  \end{center}
\end{figure}

Although percolation, i.e. 
$W>0$, does not seem to have an influence on scaling and thereby on the 
result of the multifractal analysis, there are nevertheless some effects which
have to be investigated\cite{Klesse_Metzler1}. They are very obvious if one 
looks at the wave functions themselves (see Fig.~\ref{fig:wave}), but also show
some influence in the scaling analysis.

\begin{figure}[htb]
  \begin{center}
    \leavevmode
    \epsfig{figure=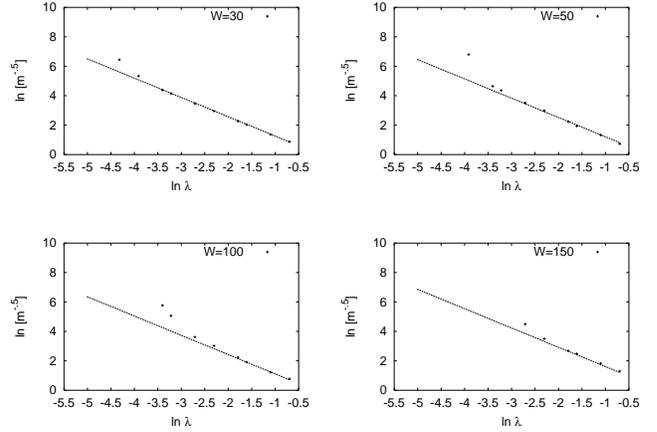,angle=270,width=.45\textwidth}
  \caption{The scaling of the average box-probabilities for different 
    values of $W$ at $q=-.5$. System size $L=250$.}
    \label{fig:scalew}
  \end{center}
\end{figure}

The eigenstates in Fig.~\ref{fig:wave} obtained in the generalized network 
model at $E=E_c=0$ show the typical self-similar shape of critical wave 
functions which led to the use of multifractal analysis on critical
systems\cite{Janssen}. 
In this context the scale invariance of the eigenfunctions is investigated
by obtaining the following quantities:
\begin{equation}
  \label{eq:boxprob}
  m(l_b)=\int_{l_b^d} d{\bf r} |\Psi({\bf r}+{\bf r}_0)|^2,
\end{equation}
which denote the probability of a particle to be in a box of linear size
 $l_b$ centered at ${\bf r}_0$, respectively. The disorder averaged $q$-moments
of $m(l_b)$, $m_q(l_b)=\langle m^q(l_b)\rangle$, scale over a wide range of 
box sizes with definite exponents,
\begin{equation}
  \label{eq:scale}
  m_q(\lambda)\propto \lambda^{d+\tau(q)},
\end{equation}
where $\lambda=l_b/L$ is the quotient of box size $l_b$ to system size $L$.
The exponents $\tau(q)$ depend non-linearly on $q$ and characterize the 
universality class of the system. Often the single exponent 
$\alpha_0=d\tau(q)/dq(q=0)$ is used instead of the entire $\tau(q)$ spectrum,
because it describes the scaling of the typical value of the squared amplitude
with the system size, $\exp\langle\ln |\Psi|^2\rangle\propto L^{-\alpha_0}$.
The results of numerical investigations of these exponent show that they do
not depend on the size of $W$ (i.e. $W/E_t$, we will set $E_t=1$ in the 
following)\cite{Klesse_Metzler1}. However, we can observe that the range of
box sizes for which scaling behavior is seen does change.
For $W<1$ the box probabilities scale over the entire range $a < l_b < L$, where
$a$ is the lattice constant of the network.  
If we increase $W$, scaling deteriorates for small box sizes,
i.e. small $\lambda$'s.
In Fig.~\ref{fig:scalew} we show the 
$\ln [m^q(l_b)]$--$\ln\lambda$-plots for four different values
of $W$ at $q=-0.5$. We see that the deterioration of scaling shows a 
significant dependence on $W$ indicating that the valid range of lengths
for scaling follows a law of the kind
\begin{equation}
  \label{range}
  a\left(\frac{W}{E_t}\right)<l_b<L,
\end{equation}
where $a(x)$ is a function we have to determine.

\begin{figure}[htbp]
  \begin{center}
    \leavevmode
    \epsfig{figure=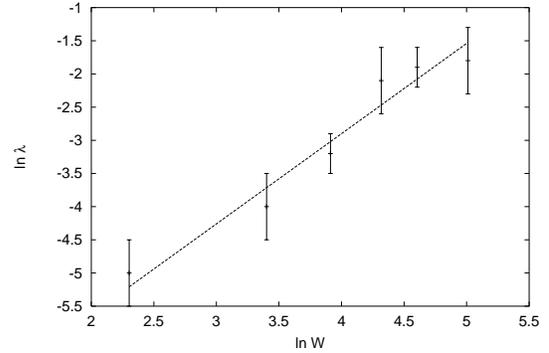,angle=270,width=.4\textwidth}
  \caption{The logarithms of the $\lambda$'s at
      which scaling deteriorates
      plotted against $\ln W$. As a result for the linear fit we get a slope
      of $1.36\pm 0.06$}
    \label{fig:lamoW}
  \end{center}
\end{figure}

If we take the values\cite{footnote}
of $\ln\lambda$ for each $W$ at which the linear
approximation becomes invalid and plot them against $\ln W$, a linear
dependence becomes apparent  (see Fig.~\ref{fig:lamoW}). A linear fit of the
data yields 
a slope of $X=1.36\pm 0.06$. Consequently, the minimal length
where scaling can be seen follows the following relation
\begin{equation}
  \label{l0}
  a \propto \left(\frac{W}{E_t}\right)^X.
\end{equation}
Another effect of the increase of $W$ are the increasing sample to sample
fluctuations and the rapid increase of the statistical error in the 
$f(\alpha)$ data, especially for large values of $|q|$. This all seems
to be accompanied by a visible change of the characteristics of the
wave function, as seen in Fig.~\ref{fig:wave}. It is obvious that 
for $W=150$ the wave function shows large areas, where the square amplitude
changes only slightly, whereas for $W=0$ such areas of constant amplitude
are much smaller. In fact one can see a steady increase of the size 
of these areas if one gradually increases $W$. 

If we take all these effects into account, we have to conclude that,
apparently, the introduction of variable saddle point energies, i.e. $W>0$, 
leads to a new length scale in our system.

In order to understand how classical percolation can alter the shape of
the quantum mechanical wave functions and influence their scaling behavior
we will repeat the discussion published previously in literature
\cite{Polyakov,Klesse_Metzler1,Metzler_PhD}.
\begin{figure}[htbp]
  \begin{center}
    \leavevmode
    \epsfig{figure=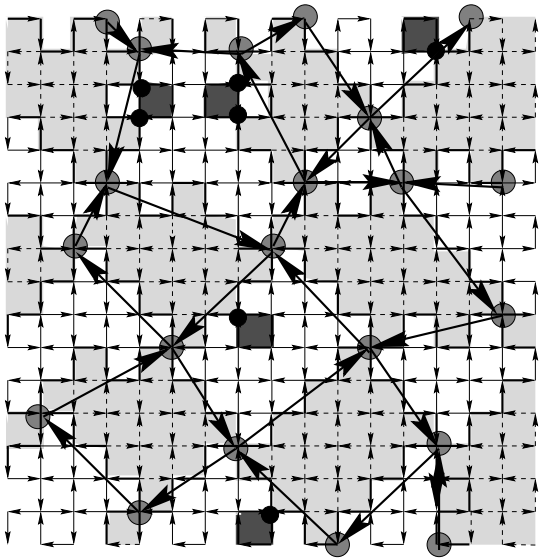,angle=270,width=.4\textwidth}
  \caption{Along the classical percolation path where the saddle point 
    energies are large, $|u_k|>E_t$, the amplitude of the
    wave functions remain mainly constant. The resulting {\em
      critical} clusters form the links of the rescaled network.}
    \label{fig:perc}
  \end{center}
\end{figure}
Take a look at those saddle points whose energy $u_k$ differs more 
than $E_t$ from the electron energy $E=0$. Of course, such saddle points 
occur only for $W>E_t$. The transmission coefficients $T_k$ at these 
saddle points are exponentially small and the amplitude
of an incoming link is essentially transmitted to a single outgoing link
(classical behavior). Therefore, on a path avoiding saddle points with
energy $|u_k|<E_t$ the amplitude of the wave function remains mainly 
constant. This leads to the formation of clusters where the wave function
amplitude changes only slightly (see Fig.~\ref{fig:perc}). 
The size of these clusters is obviously connected to classical percolation.
The cluster correlation length $\xi_p$ of clusters connected by saddle points
whose energy is smaller than $E_t$ scales according to scaling theory
(see e.g. refs.~\citen{Isichenko92:961}and \citen{Stauffer85:0}) as
\begin{equation}
  \label{dnup}
  \xi_p\propto \left(\frac{E_t}{W}\right)^{-\nu_p}.
\end{equation}
This is the size of the largest clusters connected by those saddle points.
If we consider that for smaller clusters, which connect via these 
saddle points to other large or small clusters, the effect in terms of
the change of wave function amplitude along the percolation path will be 
negligible, we see that
the relevant cluster size is of order $\xi_p$. Accordingly, not all
saddle points with $u_k<E_t$ are relevant for the quantum mechanical effects,
but only those {\em critical} saddle points that connect {\em critical} 
clusters and the scaling 
behavior of those clusters is given by (\ref{dnup}). 
If we identify the clusters 
with the links of a new network and the saddle points with its nodes, we
obtain a rescaled network model which has a renormalized lattice 
constant $a'=a(W/E_t)$ and a new $W'=E_t$. 
In this way, we can treat network models  with arbitrarily large finite $W$ by 
mapping it on the original Chalker--Coddington model\cite{Klesse_Metzler1}.
Thus, we see that the influence of percolation is not on the overall 
scaling behavior but only on the range of the scaling. Its effects are
mostly finite size effects which arise due to the decrease of the 
effective system size when we increase $W$.

One very strong finite size effect is the shift of the critical energy $E_c$
when $W$ is increased. This shift is not systematic but depends on
the individual realization of the network. Due to the increasing range 
of the saddle point energies when $W$ increases the average value of
saddle points energy for a single system is no longer $\langle
u\rangle=0$. Since the number of saddle points depends on the system
size $L$, the number of random saddle point energies drawn from the
square distribution may not be sufficient to guarantee that the
deviation from the average is small when the $W$ approaches the system 
size causing the critical point to shift to $E_c=-\langle
u\rangle$ and consequently leading to strong fluctuations in the
results for critical quantities. This suggests that any numerical  
investigations should be confined to values of $W$ that obey $W<L$.

\section{Spectral Properties}
The next step in our investigation is to see what happens to the spectral
properties of the system when $W$ is increased. In this case we are looking
at the level spacing distribution $P(s)$ and at the level number variance
$\Sigma_2(N)=\langle (n-\langle n\rangle)^2\rangle$ of energy
intervals containing on average $N=\langle n\rangle$ levels.

At the critical point $P(s)$ has a unique shape\cite{Metzler_Varga} which 
lies between the Poissonian behavior in the localized regime, 
$P(s)=\exp(-s)$, and that of the 
Gaussian unitary ensemble (GUE)-like in the metallic regime,
$P(s) =\frac{32}{\pi^2} s^2\exp\left(-\frac{4}{\pi}s^2\right)$, given by
random matrix theory(RMT)\cite{rmt}. For small $s$ the level repulsion
leads to a GUE-like behavior ($P(s)\propto s^2$), whereas for larger $s$
the tail of the distribution behaves like $P(s)\propto 
\exp\left( -\kappa s\right)$, where $\kappa<1$.
This mixture between GUE and Poisson is due to the multifractal structure
of the wave functions which are neither homogeneously smeared out over the 
entire system, as in the metallic regime, nor strongly localized. 
The direct connection between level distribution and multifractal wave
function was found by Chalker et al. \cite{CKL} when they derived for the
compressibility
$\chi=\lim_{N\rightarrow\infty}\lim_{L\rightarrow\infty}d\Sigma_2/dN$
at the critical point
that $\chi=\frac{d-D_2}{2d}$, where $d$ is the spatial
and $D_2$ the fractal (correlation) dimension of the wave function.
This equation could be verified numerically for the Chalker--Coddington
network \cite{Klesse_Metzler2} and shows that we are again between the
metallic ($\chi=0$) and the localized ($\chi=1$) behavior. 
If our arguments concerning the influence of classical percolation are valid,
we should obtain the same kind of results for the critical spectra regardless
of the value of $W$. On the other hand, we have seen that increasing $W$
causes finite size effects to appear faster which should also be visible
in the spectral properties.

\begin{figure}[htbp]
  \begin{center}
    \leavevmode
    \epsfig{figure=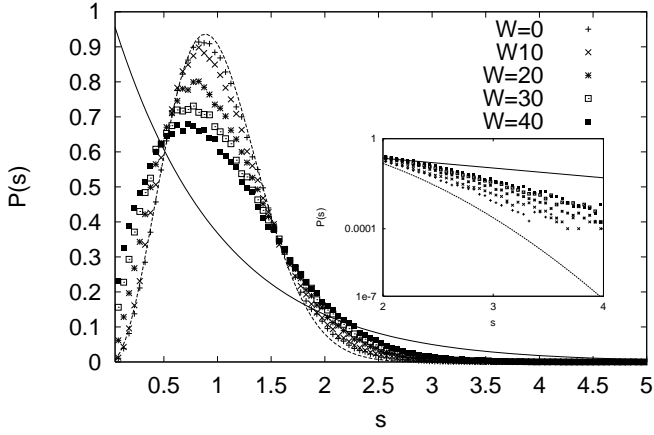,angle=270,width=.5\textwidth}
  \caption{The $P(s)$ spectrum for $W=0,10,20,30,40$. The smooth
    curves show the Poisson and GUE distributions. The inset shows the 
    same on a logarithmic scale.}
    \label{fig:pos}
  \end{center}
\end{figure}

We start by taking a look at the $P(s)$ distribution for different
values of $W$. In Fig.~\ref{fig:pos} we can see $P(s)$ for $W=0$ to $W=40$.
In this case we chose the system size $L=80$, but we obtain the same
$P(s)$ for other system sizes. The distribution is still system size
independent at least for the range of system sizes we are able to
investigate. This rules out finite size effects which could be
expected due to the decrease of effective system size 
observed for the wave functions. Nevertheless, the distribution seems
to  become more and
more  Poissonian for larger
$W$. This effect has already been seen by Ono et al. \cite{OOK} who
used the Landau model for the quantum Hall effect. In this case an increase 
of the potential's correlation length, which is according to
eq.~(\ref{eq:tunnel}) comparable to an increase of $W$, led to a similar
result. In their publication Ono et al. fitted the $P(s)$ distribution to  
the formula:
\begin{equation}
  \label{eq:fit}
  P(s)=As^\beta\exp(-Bs^\alpha),
\end{equation}
where $A$ and $B$ are determined by the normalization conditions
$\langle s\rangle=\langle 1\rangle=1$ and $\alpha$ and $\beta$ are the
fitting parameters. This way of fitting was prompted by the results of
Kravtsov et al.\cite{KLAA1,KLAA2}, which predicted that
$\beta=2$ as for the Gaussian unitary ensemble (GUE), where
$P(s)=\frac{32}{\pi^2}s^2\exp(-4/\pi s^2)$, and $\alpha=1+1/(d\nu)\approx
1.21$. Although, the numerical values for $\alpha$ do not fit that
prediction \cite{OOK,Metzler_Varga} we can still use eq.~(\ref{eq:fit})
to compare our data with that in ref.~\citen{OOK}.
The result of such a fit to the distributions in Fig.~\ref{fig:pos} is 
shown in Table~\ref{tab:albeta}.
\begin{table}[h]
  \begin{tabular}{@{\hspace{\tabcolsep}\extracolsep{\fill}}rrrrr}
    \hline
    $W$&$ \alpha $&&$\beta$&\\ 
    \hline 
    $0$&$1.54$&$\pm 0.02$&$2.03$&$\pm 0.03$\\
    $10$&$1.71$&$\pm0.04$&$1.96$&$\pm 0.05$\\
    $20$&$1.79$&$\pm 0.02$&$1.30$&$\pm 0.02$\\
    $30$&$1.94$&$\pm 0.03$&$0.85$&$\pm 0.01$\\
    $40$&$2.04$&$\pm 0.03$&$0.59$&$\pm 0.01$\\
    \hline
    $l_c$&1.52&$\pm 1.3\times 10^{-4}$&1.95&$\pm 7.3\times 10^{-4}$\\
    $2l_C$&1.98&$\pm 5.7\times 10^{-5}$&1.03&$\pm 1.6\times 10^{-3}$\\
    \hline
  \end{tabular}
  \caption{The results for the fit parameters $\alpha$ and $\beta$
    for different values of $W$. The last two entries give the
    values for different potential ranges determined by Ono et al.}
  \label{tab:albeta}
\end{table}
We can see that $W\approx 30$ corresponds to the results obtained 
by Ono et al. for a
potential correlation length of $2l_c$, whereas a correlation length of
$l_c$ did not produce significant changes from the uncorrelated
results (i.e. $W=0$). We would expect that,
because the relevant microscopic length in the Landau model is $l_c$
so that the potential correlation length has to be larger to show an
appreciable effect. This corresponds to the fact that $W\approx E_t$
also has no effect since the tunnel energy $E_t$ sets the energy scale 
for our system and $W$ has to be larger in order to obtain {\em
  classical} saddle points.

For both models it is very clear that the level repulsion for small
values of $s$ decreases with increasing $W$ and correlation length,
respectively. This is indicated by the increase of $P(s=0)$ and the
decrease of the fit parameter $\beta$. 

\begin{figure}[htbp]
  \begin{center}
    \leavevmode
    \epsfig{figure=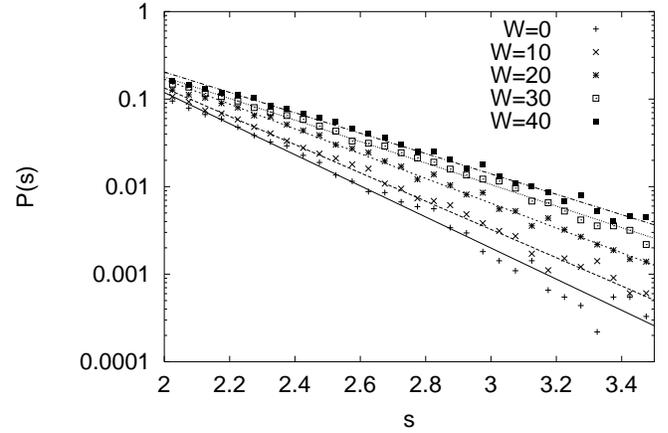,angle=270,width=.5\textwidth}
  \caption{The tail of the $P(s)$ spectrum for $W=0,10,20,30,40$. The smooth
      curves are the linear fits for respective curves.}
    \label{fig:tail}
  \end{center}
\end{figure}

\begin{figure}[htbp]
  \begin{center}
    \leavevmode
    \epsfig{figure=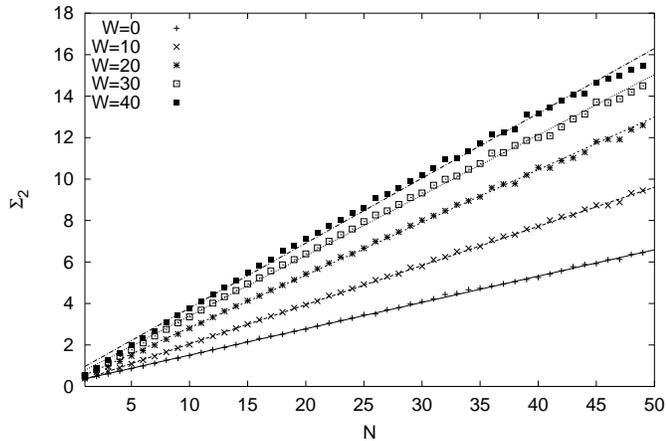,angle=270,width=.5\textwidth}
  \caption{The function $\Sigma(N)$ for different values of $W$.}
    \label{fig:chi}
  \end{center}
\end{figure}

In ref.~\citen{Metzler_Varga} it was found that the tail of $P(s)$ shows an 
exponential decay of the form $P(s)\propto \exp(-\kappa s)$, which was 
predicted by Altshuler et al. \cite{AZKS} with $\kappa=1/(2\chi)$ and
confirmed for QHE systems by the numerical results in
refs.~\citen{Metzler_Varga} and \citen{BS}. 
In Fig.~\ref{fig:tail} we show the behavior of the tail region of
$P(s)$ for different $W$. Although the tail seems to remain an
exponential the factor $\kappa$ decreases with increasing $W$ getting
closer to the Poissonian behavior, where $\kappa=1$.  
This would indicate that the compressibility $\chi$ also changes with
$W$. In Fig.~\ref{fig:chi} we show the level number variance
$\Sigma_2(N)$ from which we obtain $\chi$ by determining its slope.
We can see that the slope increases with increasing $W$ which would
correspond to the decrease of $\kappa$, but a look at
Table~\ref{tab:kapchi} shows that the relation $\kappa=1/(2\chi)$
\begin{table}[h]
  \begin{center}
    \begin{tabular}{@{\hspace{\tabcolsep}\extracolsep{\fill}}rrrrrrr}
      \hline
      $W$&$\kappa$&&$\chi$&&$1/(2\chi)$&\\
      \hline
      $0$&$4.09$&$\pm 0.10$&$0.127$&$\pm 0.002$&$3.94$&$\pm 0.06$\\
      $10$&$3.72$&$\pm 0.06$&$0.190$&$\pm 0.003$&$2.63$&$\pm 0.04$\\
      $20$&$3.26$&$\pm 0.05$&$0.254$&$\pm 0.001$&$1.96$&$\pm 0.01$\\
      $30$&$2.83$&$\pm 0.04$&$0.291$&$\pm 0.002$&$1.71$&$\pm 0.01$\\
      $40$&$2.68$&$\pm 0.04$&$0.313$&$\pm 0.002$&$1.59$&$\pm 0.01$\\
      \hline
    \end{tabular}
    \caption{The results for $\kappa$ and $\chi$ for different values of $W$.}
    \label{tab:kapchi}
  \end{center}
\end{table}
no longer holds in the case $W>0$. This is also true for the relation
$\chi=\frac{d-D_2}{2d}$ connecting the compressibility $\chi$ to the
multifractal exponent $D_2$\cite{localized}. From the investigation of
the critical
wave functions we know that $D_2$ does not change with $W$. 

\begin{figure}[htbp]
  \begin{center}
    \leavevmode 
    \epsfig{figure=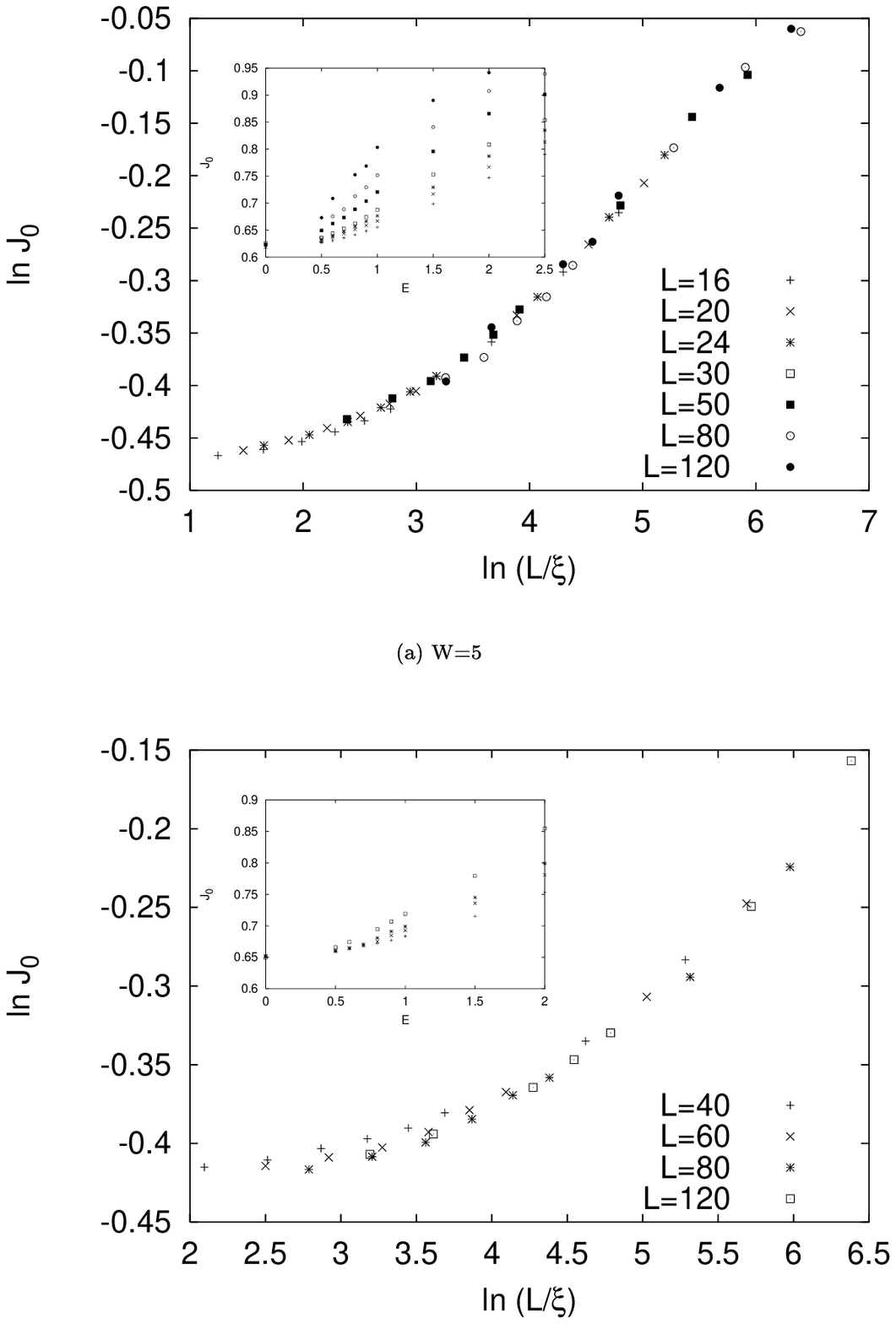,width=.5\textwidth}
  \caption{The one-branch scaling functions $J_0(L/\xi)$ for four
    different values of $W$ resulting from 
    the rescaling of the function $J_0(E,L)$ shown in the insets.}
    \label{fig:w2-15}
  \end{center}
\end{figure}

Finally, observing that increasing $W$ has a strong effect on the shape
of $P(s)$ the question arises whether a function like
\begin{equation}
  \label{eq:j0}
  J_0(L,E)=\frac{1}{2}\langle s^2\rangle_{L,E},
\end{equation}
which depends on as well as describes the shape of $P(s)$ still shows
the same one-parameter scaling for $W>0$ as for
$W=0$. The insets in Fig.~\ref{fig:w2-15} shows $J_0$ for 
$W=5,15$ as a function of $E$ for different system sizes. 
At the critical point $E_c=0$ all curves come together at a single
size independent point which varies with $W$ reflecting the dependence 
of the shape of $P(s)$ on $W$. If one-parameter scaling still holds
then a scaling function $f$ of the following form should exist:
\begin{equation}
  \label{eq:scalingf}
  J_0(E,L)=f(L/\xi),\quad \xi=\xi_0|E-E_c|^{-\nu}.
\end{equation}
We determined $f(x)$ by fitting a fourth degree polynomial in
$x=\ln(L/\xi_0)+\nu\ln E$ to our data using $\xi_0$ and $\nu$ as
fitting parameters. The values we found for $\nu$ and the critical
values of $J_0$ are given in Table~\ref{tab:nuj0}.
Although the critical values $J_0^c(W)$ vary systematically with $W$
the  scaling
exponent $\nu$ shows no systematic dependence on $W$ and is
consistent with previous results for $\nu$ obtained by various
studies\cite{nu1,nu2,nu3,nu4,nu5} within the error bars. 
\begin{table}[h]
  \begin{center}
    \begin{tabular}{@{\hspace{\tabcolsep}\extracolsep{\fill}}rrrrr}
      \hline
      $W$&$\nu$&&$J_0^c$&\\
      \hline
      $0$&$2.1$&$\pm 0.3$&$0.600$&$\pm 0.005$\\
      $2$&$2.1$&$\pm 0.2$&$0.601$&$\pm 0.01$\\
      $5$&$2.2$&$\pm 0.1$&$0.604$&$\pm 0.01$\\
      $10$&$2.1$&$\pm 0.3$&$0.611$&$\pm 0.01$\\
      $15$&$2.3$&$\pm 0.1$&$0.623$&$\pm 0.01$\\
      $30$&$2.3$&$\pm 0.4$&$0.664$&$\pm 0.01$\\
      $40$&$2.2$&$\pm 0.3$&$0.685$&$\pm 0.01$\\
      \hline
    \end{tabular}
    \caption{The results for the scaling exponent $\nu$ and the
      critical value of $J_0$ for different values of $W$.}
    \label{tab:nuj0}
  \end{center}
\end{table}

\section{Conclusion}
In this paper we have discussed the influence of percolation effects
on critical wave functions and the critical quasienergy spectrum of the
Chalker--Coddington network model. Classical percolation effects
appear in the generalized version of the model when the range $W$ of the
saddle point energies, i.e. node energies of the network, is increased 
from $W=0$ in the original model to $W>1$. 

We found that multifractal analysis of the critical wave functions
results in the same exponents
regardless of the value of $W$. Nevertheless, we can see the effect of 
classical percolation in the appearance of a new microscopic
length scale which increases with $W$ and scales with an exponent
$X= 1.36\pm 0.06$ consistent with the exponent 
$\nu_p=4/3$ of the classical correlation length. This confirms the
connection of the correlation length to that new length scale, which
sets the minimum length for scaling and therefore reduces the
effective size of the system, thereby causing increasing fluctuations
when $W$ grows. 

In the case of the spectral statistics of the quasienergies of the
network the influence of an increasing $W$ is more profound. An
increasing shift from the critical statistics towards Poissonian
behavior is observed for all the spectral quantities we have
investigated. Although this shift does not seem to have an influence
on the scaling exponent $\nu$ it is clearly visible. Since the effect
is not changing with system size, it can not be ruled as a mere finite 
size effect resulting from the decrease of effective system size
observed for the critical wave functions. Furthermore, we observed
that the changes that come with an increase of $W$ are very similar to 
those connected with the increase of the potential correlation length
observed by Ono et al.\cite{OOK}.

It seems, although the multifractal behavior of the critical wave
functions is well understood, that more elaborate analytical 
studies are indicated to understand
the level statistics at the LD transition.

\section*{Acknowledgments}
The author would like to thank Y. Ono and I. Varga for useful discussions.
This work was supported by the Deutsche Forschungsgemeinschaft (DFG)
postdoc program. 
\begin{appendix}
  \section{Degeneracy of Eigenvalues}
\label{appendix}
The chiral structure of the Chalker-Coddington network leads to a pseudo 
degeneracy of the eigenvalues of its network operator $U$.
For each eigenvector $\Psi_\alpha$ with eigenvalue $e^{i\omega_\alpha}$ there 
exits an eigenvector $\tilde\Psi_\alpha$ with eigenvalue 
$e^{i(\omega_\alpha+\pi)}$.

In order to prove this, let us first point out the fact that the operator
$U$ maps vertical links only to horizontal links and vice versa. This
means that $U^2$ maps vertical to vertical and horizontal to horizontal links,
thereby creating two orthogonal $U^2$-invariant subspaces. 
We can therefore define
the projection operators $P_v$ and $P_h$ which will project onto the
two subspaces, respectively. 

Let $\Psi_\alpha$ be
an eigenvector  of $U$ with eigenvalue $e^{i\omega_\alpha}$. Then $\Psi_\alpha$
is also an eigenvector of $U^2$ with eigenvalue $e^{2i\omega_\alpha}$.
We can write $\Psi_\alpha$ as a linear combination of the two projections
$\Psi^v_\alpha=P_v\Psi_\alpha$ and $\Psi^h_\alpha=P_h\Psi_\alpha$:
\begin{equation}
  \label{psialph}
  \Psi_\alpha=\Psi^v_\alpha+\Psi^h_\alpha=v{\bf e}^v_\alpha+h{\bf e}^h_\alpha,
\end{equation}
where $v=|\Psi^v_\alpha|$, $h=|\Psi^h_\alpha|$,
${\bf e}^v_\alpha=v^{-1}\Psi^v_\alpha$ and 
${\bf e}^h_\alpha=h^{-1}\Psi^h_\alpha$.
Since $\Psi^v_\alpha$ and $\Psi^h_\alpha$ are orthogonal to each other
they are both also eigenvectors of $U^2$ with eigenvalue $e^{2i\omega_\alpha}$.
This means that this eigenvalue is degenerate. 

Let us now construct the 
following eigenvector of $U^2$:
$$
  \tilde\Psi_\alpha=h{\bf e}^v_\alpha-v{\bf e}^h_\alpha.
$$
$\tilde\Psi_\alpha$ is obviously orthogonal to $\Psi_\alpha$. 
In order to see what will happen if we
use $U$ on $\tilde\Psi_\alpha$, let us first use it on $\Psi_\alpha$
and keep in mind that $U{\bf e}^h_\alpha=e^{i\omega_h}{\bf e}^v_\alpha$ and 
$U{\bf e}^v_\alpha=e^{i\omega_v}{\bf e}^h_\alpha$, where 
$e^{2i\omega_v}=e^{2i\omega_h}=e^{2i\omega_\alpha}$.
\begin{equation}
  \label{Upsialph}
  U \Psi_\alpha=v e^{i\omega_v} {\bf e}^h_\alpha
  +h e^{i\omega_h} {\bf e}^v_\alpha=e^{i\omega_\alpha}\Psi_\alpha.
\end{equation}
From eqs.~(\ref{Upsialph}) and (\ref{psialph}) it follows that 
$v=h e^{i(\omega_\alpha-\omega_v)}$
and $h =v e^{i(\omega_\alpha-\omega_h)}$ and consequently
\begin{eqnarray*}
  U\tilde\Psi_\alpha&=&h e^{i\omega_v} {\bf e}^h_\alpha
-v e^{i\omega_h} {\bf e}^v_\alpha\\
&=&e^{2i(\omega_\alpha-\omega_v)}h e^{i\omega_v}{\bf e}^h_\alpha
-e^{2i(\omega_\alpha-\omega_h)}v e^{i\omega_h} {\bf e}^v_\alpha\\
&=& e^{i\omega_\alpha}\left[he^{i(\omega_\alpha-\omega_v)}{\bf e}^h_\alpha
  -e^{i(\omega_\alpha-\omega_h)}v{\bf e}^v_\alpha\right]\\
&=& -e^{i\omega_\alpha}\left[-v{\bf e}^h_\alpha+h {\bf e}^v_\alpha\right]\\
&=& -e^{i\omega_\alpha}\tilde\Psi_\alpha.
\end{eqnarray*}
This means $\tilde\Psi_\alpha$ is an eigenvector of $U$ with eigenvalue
$e^{i(\omega_\alpha+\pi)}$.

The same kind of argument leads to a further pseudo-degeneracy when we
impose double periodic boundary conditions (torus) with system size $L$ 
a multiple of 2. In that case each eigenphase $\omega_\alpha$ is accompanied 
by three other eigenphases $\omega_\alpha+\pi/2$, $\omega_\alpha+\pi$
and $\omega_\alpha+3\pi/2$. This is caused by the chiral symmetry above and
the boundary conditions which lead to the formation of two more (in this case
$U^4$) invariant subspaces within each of the $U^2$-invariant subspaces of 
vertical and horizontal lines. These subspaces consist of the sets combining 
every second vertical row or every second horizontal column, respectively.
Without the boundary conditions those subspaces mix.
\end{appendix}


\begin{thebibliography}{10}

  \def \pra#1#2#3#4{ #1: Phys.~Rev.~A {\bf #2}(#4) #3.} 
  \def \prb#1#2#3#4{ #1: Phys.~Rev.~B {\bf #2}(#4) #3.} 
  \def \prl#1#2#3#4{ #1: Phys.~Rev.~Lett. {\bf #2}(#4) #3.}

\bibitem{Chalker_Coddington} J. T. Chalker, P. D. Coddington, J. Phys. C
{\bf 21}, 2665 (1988).

\bibitem{book} 
M. Janssen, O. Viehweger, U. Fastenrath and J. Hajdu: {\it Introduction
to the Theory of the Integer Quantum Hall Effect} (VCH, Weinheim; New York; 
Basel; Cambridge; Tokyo, 1994).

\bibitem{Eastmond}
J. F. G. Eastmond, Ph. D. thesis, Oxford University (1992).


\bibitem{LWK}
\prl{D. H. Lee, Z. Wang and S. Kivelson}{70}{4130}{1993}

\bibitem{Klesse_Metzler1}
R. Klesse and M. Metzler, Europhys. Lett. {\bf 32}(1995) 229.

\bibitem{footnote}
We read the data directly from plots like in 
Fig.~\ref{fig:scalew} with error bars corresponding to the possible errors made
by doing so.

\bibitem{Polyakov}
\prl{D. G. Polyakov and M. E. Raikh}{75}{1368}{1995}

\bibitem{Fertig} 
\prb{H.~A.~Fertig}{38}{996}{1988}

\bibitem{Klesse_Metzler3}
R. Klesse and M. Metzler, to be published.

\bibitem{Klesse_PhD}
R. Klesse, Ph.D. Thesis, Universit\"at zu K\"oln, (AWOS-Verlag, Erfurt) 1996.

\bibitem{Huckestein_Klesse}
\prb{B. Huckestein and R. Klesse}{55}{R7303}{1997}

\bibitem{Klesse_Metzler2}
\prl{R. Klesse and M. Metzler}{79}{721}{1997}

\bibitem{Metzler_Varga}
M. Metzler and I. Varga, J. Phys. Soc. Jpn. {\bf 67}, 1856 (1998).

\bibitem{Metzler_PhD}
M. Metzler, Ph.D. Thesis, Universit\"at zu K\"oln 1996.

\bibitem{Janssen} M. Janssen: Int. J. Mod. Phys. {\bf 8}(1994) 943.

\bibitem{Isichenko92:961}
M.B. Isichenko, Rev. Mod. Phys. {\bf 64}, 961 (1992).

\bibitem{Stauffer85:0}
D. Stauffer:{\em {I}ntoduction to {P}ercolation {T}heory},
 {T}aylor \& {F}rancis, 1985.


\bibitem{CKL}
J. T. Chalker, V. E. Kravtsov and  I. V. Lerner: JETP Lett. {\bf 64}(1996) 386. 

\bibitem{KLAA1} \prl{V. E. Kravtsov, I. V. Lerner, B. L. Altshuler and
A. G. Aronov}{72}{888}{1994}

\bibitem{KLAA2}
A. G. Aronov, V. E. Kravtsov and I. V. Lerner: JETP Lett. {\bf 59}(1994) 39.


\bibitem{OOK} 
Y. Ono, T. Ohtsuki and  B. Kramer: J. Phys. Soc. Jpn. {\bf 65}(1996) 1734.

\bibitem{rmt} M. L. Mehta: {\em Random Matrices}, 2nd ed. (Academic
Press, New York, 1991).


\bibitem{AZKS} B. L. Altshuler, I. Kh. Zharekeshev, S. A. Kotochigova and 
B. Shklovskii: Sov. Phys.-JETP {\bf 67}(1988) 625.

\bibitem{BS}
M. Batsch and L. Schweitzer: {\it High Magnetic Fields in the Physics
of Semiconductors II: Proceedings of the International Conference, 
W\"urzburg 1996}, edited by G. Landwehr and W. Ossau (World Scientific
Publishers Co., Singapure, 1997), pp. 47--50.

\bibitem{localized}
Both relations do not hold for the localized regime where $\chi=1\neq 1/2$. 
This is to be expected since the assumptions made for their derivation 
are not valid in the localized regime.(See also ref.~\citen{CKL}).

\bibitem{nu1}
\prl{H. Aoki and T. Ando}{54}{831}{1985}
\bibitem{nu2}
\prl{B. Huckestein and B. Kramer}{64}{1473}{1990}
\bibitem{nu3}
\prl{Y. Huo and R. N. Bhatt}{68}{1375}{1992}
\bibitem{nu4}
\prb{Dongzi Liu and S. Das Sarma}{49}{2677}{1994}
\bibitem{nu5}
T. Ohtsuki and Y. Ono: J. Phys. Soc. Jpn. {\bf 64}(1995) 4088. 
\end{thebibliography}
\end{document}